# Gate Voltage-Controlled Magnetic Anisotropy Effect on Pt-Porphyrin functionalized single-layer graphene

Ambika Shanker Shukla[1,2]*, Abhishek Erram[1], Heston Alfred Mendonca[1], Deepak Kumar[1,3], Akanksha Chouhan[1], and Ashwin A. Tulapurkar[1]*

[1]Solid State Devices Group, Department of Electrical Engineering, Indian Institute of Technology- Bombay, Mumbai 400076, India

[2]Seagate Technology Northern Ireland United Kingdom BT48 0BF

[3]WPI Advanced Institute for Materials Research, Tohoku University, Katahira 2-1-1, Sendai 980-8577, Japan

*We report a novel approach to engineering large voltage-controlled magnetic anisotropy (VCMA) and enhanced spin-orbit coupling (SOC) at the interface of single-layer graphene (SLG) and NiFe (Py) through non-covalent functionalization with Platinum (II) 5,10,15,20-tetraphenyl porphyrin (Pt-porphyrin). Using chemical vapor deposition (CVD) grown SLG, we demonstrate that Pt-porphyrin functionalization significantly increases the SOC and enables robust voltage modulation of interfacial magnetic anisotropy, as confirmed by spin-torque ferromagnetic resonance (ST-FMR) measurements. A substantial VCMA coefficient of 375.6 ($fJ·V^{−1}·m^{−1}$) is achieved, accompanied by an order-of-magnitude enhancement in spin torque efficiency ($θ_{sh}$) compared to pristine SLG. The resonance field exhibits a clear, reversible shift under applied gate voltage, confirming robust electric-field modulation of interfacial magnetic anisotropy. Raman spectroscopy and X-ray photoelectron spectroscopy (XPS) confirm the structural integrity and effective charge transfer at the functionalized interface. Electrical characterization of back-gated graphene field-effect transistors (GFETs) further reveals tunable electronic properties upon functionalization. Our results establish functionalized graphene/ferromagnet interfaces as a promising platform for low-power, voltage-controlled spintronic devices, paving the way for scalable, energy-efficient memory and logic technologies*

## Introduction

The VCMA effect manipulates surface magnetic anisotropy energy when subjected to an electric field [1][2], garnering significant attention for its ability to enable fast and energy-efficient magnetization switching in spintronic devices [3]. However, the search for materials having a large VCMA coefficient is a contemporary research interest among the spintronics community to develop non-volatile (NV) memory devices with standalone VCMA switching [4][5][6]. In this search for materials, a class of 2D materials having atomic thinness of the layers leads to strong electrostatic gating effects, which can render a large VCMA effect [7]. It's been two decades since the discovery of graphene [8] [9] and it turns out to be advantageous for spintronics due to its substantial spin diffusion length [10] However, due to the low spin-orbit coupling in graphene, the conversion efficiency of charge to spin is low, which hinders application of graphene to switching of memory devices.

Hence, the enhancement of SOC in graphene is presently a significant area of interest within the academic community. Many approaches are employed to boost the SOC, including the decoration of graphene with ad-atoms such as hydrogen or organic aromatic molecules, as well as the hydrogen silsesquioxane irradiation technique, which leads to increased SOC [11].

In this article, we focus on an experimental study involving non-covalently decorated metalloporphyrin at an SLG/Py interface, highlighting enhanced spin-orbit coupling and a notable VCMA effect. We also considered the fact that the VCMA coefficient can be enhanced with interface engineering by the insertion of heavy metals such as Iridium [12], even conventional Spin-Orbit Torque (SOT) devices consist of a bilayer material stack of ferromagnetic and heavy metals like tungsten and platinum [13][14]. Replacing these costly materials with affordable 2D alternatives like SLG will not only offer a significant cost reduction but also minimize eddy current losses, a vital factor for high-frequency spintronics applications. We believe our approach has the potential to revolutionize memory systems by enabling the development of more energy-efficient and cost-effective spintronic-based devices, driving innovation in the memory industry [15].

**Results and Discussion**

The first experiment we did to characterize the pristine and functionalized SLG was RAMAN spectroscopy, where we obtained spectra on samples from 1000 cm$^{-1}$-3000 cm$^{-1}$. For pristine SLG, the $D$, $G$, and 2$D$ peaks were prominent in the spectrum, at 1360 cm$^{-1}$, 1560 cm$^{-1}$, and 2680 cm$^{-1}$, respectively. The $G$ mode corresponds to the in-plane vibrational mode of sp² carbon atoms and the $E_{2g}$ phonon at the Brillouin zone center, and the $D$ mode is due to the breathing modes of sp$^2$ atoms and requires a defect for its activation. The most prominent feature of graphene is the second-order of the $D$ mode, the 2$D$ mode. 2$D$ mode is a signature of the second-order phonon and is always present even if there is no $D$ mode. 2$D$-band position can provide insights into the number of graphene layers [16], defect density [17], or electronic changes induced by porphyrin functionalization [18]. We fitted a single Lorentzian curve on a 2$D$ mode peak and verified that the transferred graphene is a monolayer.

The FWHM for the pristine SLG for the 2$D$ band and $G$-band are 28.76 cm$^{-1}$ and 12.1 cm$^{-1}$, respectively. We calculated an $I_{2D}/I_G$ of 1.24 for pristine SLG. We functionalized pristine SLG with Pt-porphyrin with 3 different molarities named as $M$, $M$/4, and $M$/10, respectively. The process of preparing different molarities is explained in the Method section. We calculated an increase in the FWHM for both the 2$D$ and $G$-bands as depicted in Table-1.

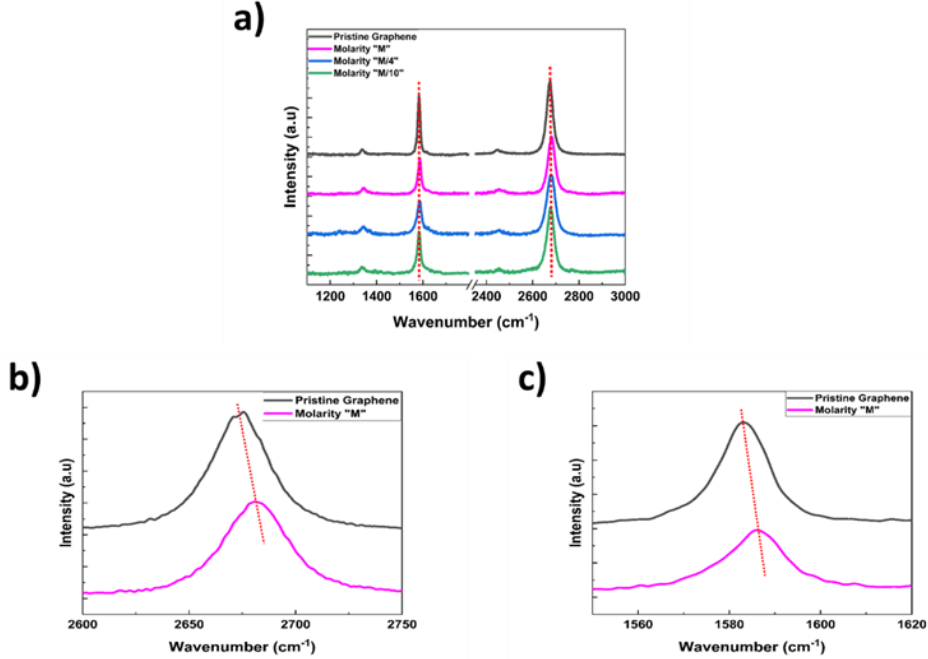

Figure 1. (a) RAMAN spectrum of SLG and functionalized SLG with *M*, *M*/4, and *M*/10 Molarity of Pt-Porphyrin solution. (b) and (c) Magnified spectra of *G*- and 2*D*-band, respectively, showing a red shift in peak position upon functionalization of SLG with *M* molarity Pt-Porphyrin

We also calculated the enhancement in the $I_{2D}/I_G$ ratio upon functionalizing pristine SLG with different molarity samples, and the calculated values are presented in Table 1.

The enhancement of the 2*D* mode upon functionalization with Pt-porphyrin signifies an interaction between the π-electrons of graphene and the molecular orbitals of Pt-porphyrin, modifying the electron-phonon coupling within the graphene sheet. This will lead to charge transfer between graphene and the Pt-Porphyrin molecules and can cause phonon modes to modulate, enhancing the 2*D* mode. We observed no change in the *D*-band FWHM and Intensity values after non-covalent functionalization, confirming minimal low defect density and structural damage to SLG.

| Molarity | $I_{2D}/I_G$ | 2D peak FWHM |
|---|---|---|
| Pristine | 1.21 | 28.24 |
| *M* | 1.52 | 43.61 |
| *M*/4 | 1.59 | 36.25 |
| *M*/10 | 1.74 | 34.26 |

Table 1. Analyzed data for $I_{2D}/I_G$ intensity peak ratio for *M*, *M*/4, and *M*/10 molarity samples and Full Width Half Maximum (FWHM) values for the 2*D* Peak.

We employed the Cancado equation [19] [20] (Eq. (1)) , to calculate the crystal size of SLG from the Raman data, where $L_a$ corresponds to the crystallite size, $\lambda_{laser}$ represents the wavelength of the excitation laser, $\lambda_{laser}$ = 532 nm, $I_D/I_G$ is the intensity ratio of the *D* and *G* peaks, $I_D/I_G$ is 2.2 in our case and 2.4e-10 is the proportionality constant between $I_D/I_G$ and $L_a$.

$$L_a = (2.4E-10) * (\lambda_{laser})^4 (\frac{I_D}{I_G})^{-1} (nm) \quad (1)$$

The average nanocrystallite size, $L_a$, is 8.73 nm, measured for the SLG on the oxidized substrate surface. Analysis of RAMAN data indicated that our chosen functionalization approach consistently and uniformly grafted the SLG with Pt-Porphyrin, demonstrating good repeatability.

We then performed XPS characterization to study further the elemental composition of the functionalized surface and the chemical bonding between pristine graphene and Pt-porphyrin. We prepared two samples for the XPS experiment: (i) Transferred Pristine SLG graphene onto the oxidized silicon substrate to obtain a reference to carbon peaks. (ii) A stack of Pt-Porphyrin functionalized SLG/Ni (2nm)/SiO$_2$ (2nm). We performed depth-resolved XPS characterization to study the presence of non-covalent bonds for Pt-Porphyrin molecules and pristine graphene [19]. We recorded a survey spectrum of the second sample as shown in Fig. 3 and observed the presence of the following elements Pt, Ni, Porphyrin ring Nitrogen, and Carbon atoms. We observed a shift of 0.3(eV) in Binding Energies (B.E) for carbon C1S peaks after functionalization and occurrences of new higher energy bonds at 289.27(eV) and at 285.96 (eV), representing C-N-H bonding with Nitrogen ring of Porphyrin and C-OH bonds representing Phenol bonding with carbon. The appearance of these bonds gives a signature of charge transfer between SLG and Pt-Porphyrin and a non-covalent bonding between them.

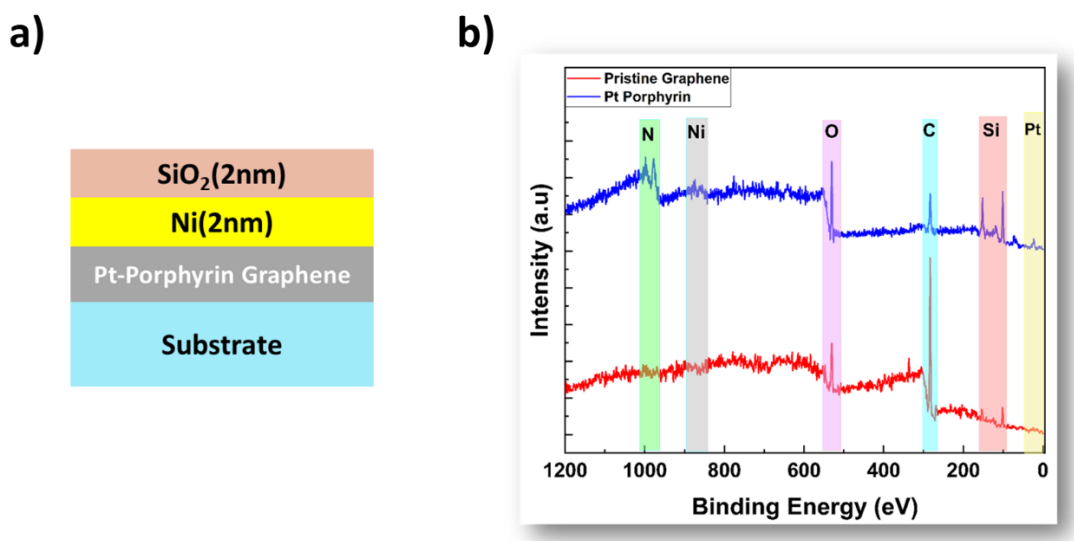

Figure 2  Depth resolved survey graph of a pristine single-layer graphene and a stack of Pt-porphyrin functionalized SLG/Ni (2nm)/SiO$_2$ (2nm) stack. For pristine graphene, we can only observe prominent carbon peaks along with oxygen and silicon substrate peaks, whereas in the Pt-Porphyrin functionalized sample we can observe the presence of Nitrogen peaks, which are the main constituents of the porphyrin ring. We can also observe distinct Platinum (Pt) and Nickel (Ni) peaks. We de-convoluted the XPS peaks

using the Gaussian function and observed sub-peaks at higher energy, which signifies non-covalent bonds of a carbon atom of pristine graphene and Pt-Porphyrin.

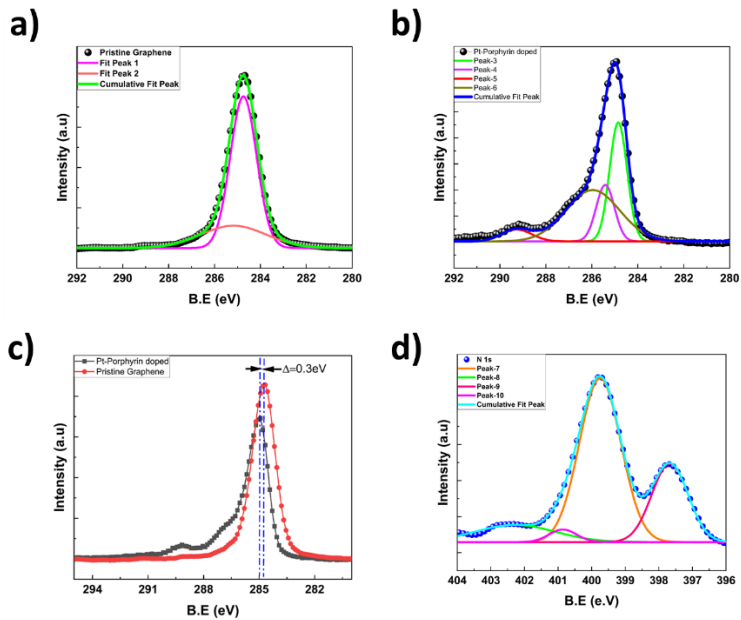

Figure 3 Carbon spectrum for a)Pristine and b) for Pt-Porphyrin functionalized samples having de-convoluted Peak-1( C1s) and Peak-2(C=C), Peak-4(C-N-H), Peak-5(C-OH) c) We observe a 0.3eV shift in binding energy in carbon peak after functionalization. d) We plotted the Nitrogen peak with de-convoluted peaks Peak-7 (Pyridinic Nitrogen) and Peak-8 (Pyrrolic Nitrogen) and observed along with higher energy bonds representing the hybridization of graphene carbon atoms and porphyrin molecule.

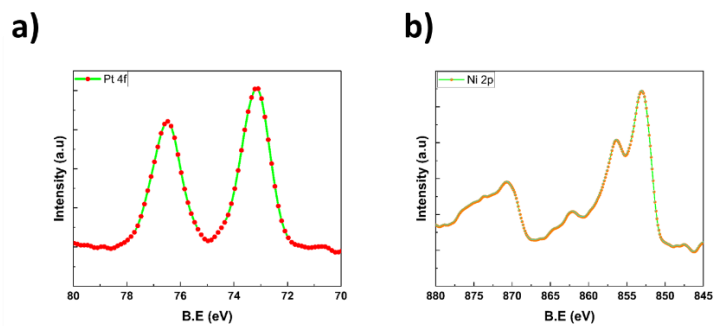

Figure 4. (a) Peaks at 73.15 (eV) and 76.46 (eV) confirm the presence of Pt atom on the functionalized surface. (b) We observed a prominent Ni peak at 853.07 (eV) and 870.7 (eV) in the prepared sample confirming the presence of Ni atoms.

Figures 4 (a) and (b) display the Pt and Ni peaks, thereby confirming their presence on the functionalized SLG surface. The detection of Ni peaks validates the thin film deposition approach, while the identification of Pt peaks indicates the preservation of metalloporphyrin on the surface of SLG, demonstrating no adverse effects on the thin film deposition, hence affirming the efficacy of the grazing angle deposition methodology [22].

To study the electrical behavior of Pt-Porphyrin we fabricated back-gated Graphene-Field Effect Transistors (GFET) devices as shown in Fig. 6(b) [23]. After grafting the surface of SLG with Pt-Porphyrin and analyzing DC characterization results, we observed a blue shift in the Dirac point with a decrease in graphene channel resistance. We can conclude that SLG works as an electron donor and Pt-Porphyrin molecule works as an electron acceptor [18]. For the graphene channel with length $L$=20um and width $W$=10um, we noted a Dirac point at -29.4V for pristine SLG, and with Pt-Porphyrin functionalization we measured a shift in value, and a new value of Dirac point was noted at -7.6V

| Element Name | Peaks | Bonds | B.E(eV) before doping | B.E(eV) after doping | Shift in B.E (eV) | FWHM after doping | FWHM before doping |
|---|---|---|---|---|---|---|---|
| C 1s | Peak-1 | C 1s | 284.74 | 284.84 | | 0.895 | 1.35 |
| | Peak-2 | C=C | 285.17 | 285.41 | 0.3 | 0.93 | 3.34 |
| | Peak-3 | C-N-H | NA | 289.27 | | 1.54 | NA |
| | Peak-4 | C-OH | NA | 285.96 | | 2.75 | NA |

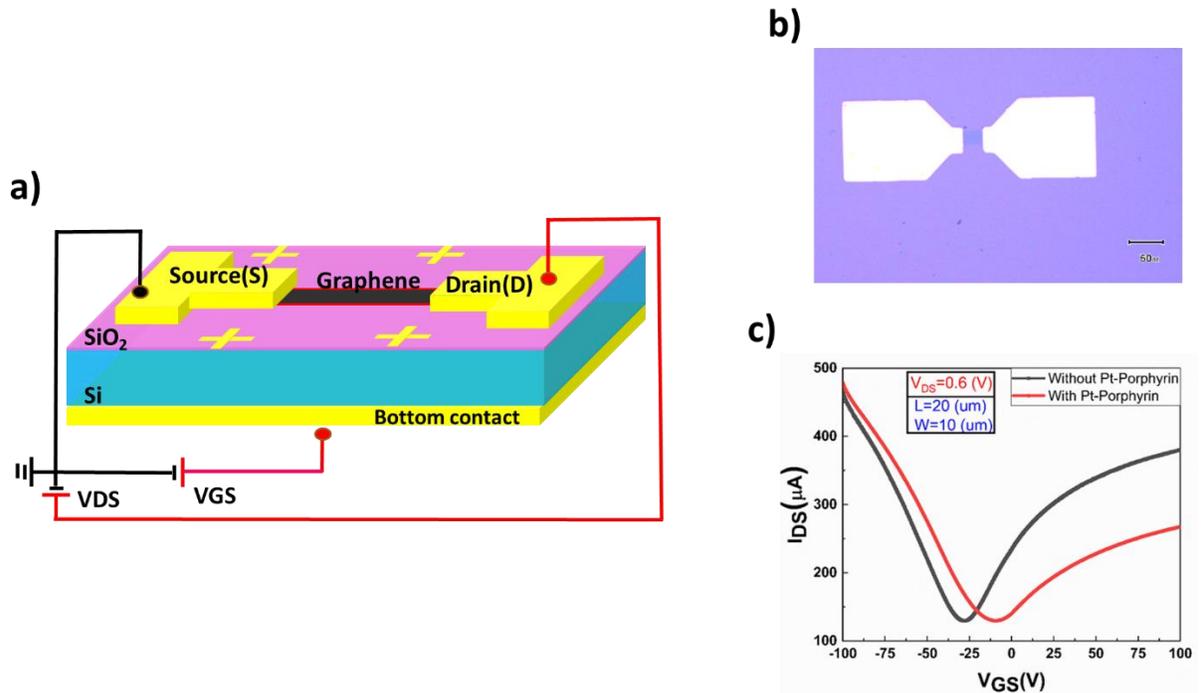

Figure 5. DC characterization of GFET device: (a) Schematic representation of GFET devices, graphene is the conductance channel between two electrical contacts of source(S) and drain (D). $V_{DS}$ is the drain voltage and $V_{GS}$ is the gate voltage applied to the source, respectively. We swept $V_{GS}$ from ±100V, and measurements were taken at individual single-valued $V_{DS}$. We characterized devices for $V_{DS}$ from ±1V at a step of ±0.1V (b) Top view Optical Image of fabricated GFET devices c) DC characterization results, measured drain to source current($I_{DS}$) as a function of gate to source voltage($V_{GS}$), and plotted $I_{DS}$ vs $V_{GS}$ for pristine SLG (Red) and Pt-Porphyrin functionalized SLG(Black). A positive shift in the Dirac point after functionalizing the SLG can be observed from the graph. SLG had a negative Dirac voltage and upon functionalization, it had a blue shift which signifies that Pt-Porphyrin acts as an electron acceptor and SLG as a donor.

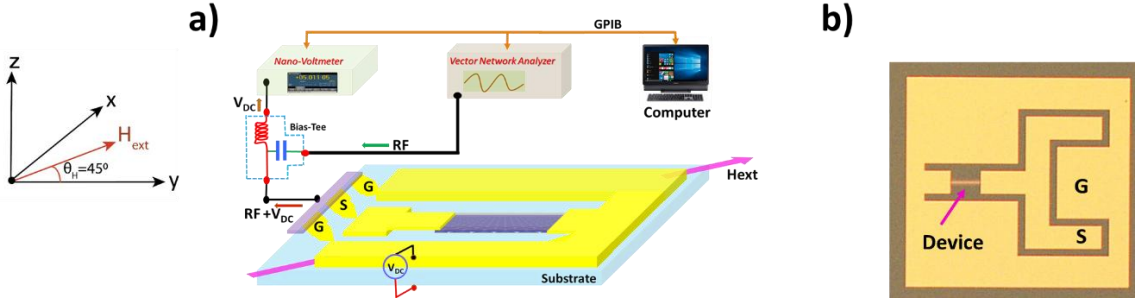

Figure 6. (a) Schematic of device used for ST-FMR measurements. We applied an RF signal using a Vector Network Analyzer (VNA) which works as an RF source, and measured the generated $V_{dc}$ at the same time with the help of Bias-Tee. We applied DC voltage ($+V_{DC}$) at the back gate using a Keithley-2602B DC source and $-V_{DC}$ at the ground pad of the device. (b) Optical Image of the fabricated device having a rectangular strip (as marked by an arrow) of the stack connected through a coplanar waveguide having a characteristic impedance of 50Ω.

We fabricated devices as shown in Fig.6 (b), a detailed process flow is discussed in the Device Fabrication and Characterization section later in this report. Py is magnetized in-plane, i.e. along the y-axis and has an out-of-plane hard axis, i.e., along the z-axis. As discussed in the Device Fabrication and Characterization section, we fabricated devices to record the ST-FMR spectrum at every single RF frequency ($f_{RF}$) starting between 4 to 7 GHz. We couldn't go beyond 7GHz as the signal strength becomes weak to detect. In our device, upon passing $I_{RF, charge}$ current Pt-Porphyrin functionalized SLG with enhanced SOC converts $I_{RF, charge}$ current into $I_{RF, spin}$ current via spin-Hall effect, and these injected spins will get absorbed by the Py layer. These absorbed spins with orientation perpendicular to both charge current and spin current flow will exert torque on magnetization ($M$) of the Py layer and it undergoes small oscillations because of magnetization oscillation, Anisotropic magneto-resistance (AMR) will also oscillate. A homodyne mixture of AMR resistance and $I_{RF}$ produces a large DC voltage ($V_{DC}$) at resonance because of the spin torque diode effect [25].

To demonstrate the VCMA effect, we apply an external DC voltage and record the ST-FMR spectrum. Applied DC voltage will generate an Electric field ($\vec{E} = \frac{V_{ox}}{t_{ox}}$) across the oxide layer, and this will modulates the perpendicular magnetic anisotropy at the interface and will eventually modulate the resonance field of the ST-FMR spectrum. We recorded the spectrum under different back gate potentials between ±25V in a step of 10V.

We analyzed the $V_{dc}$ signal by fitting symmetric and anti-symmetric Lorentzians as given by Eq.(2), where $V_{sym}$ and $V_{Asym}$ are symmetric and asymmetric components, respectively, $\Delta H$ is the line-width of the resonance signal, $H_{ext}$ is an external applied magnetic field, and the $H_{reso}$ is the resonance field of the resonance spectrum.

$$V_{dc} = V_{sym}\left[\frac{\Delta H^2}{4*(H_{ext}-H_{reso})^2 + \Delta H^2}\right] + V_{Asym}\left[\frac{4*\Delta H*(H_{ext}-H_{reso})}{4*(H_{ext}-H_{reso})^2 + \Delta H^2}\right] \qquad (2)$$

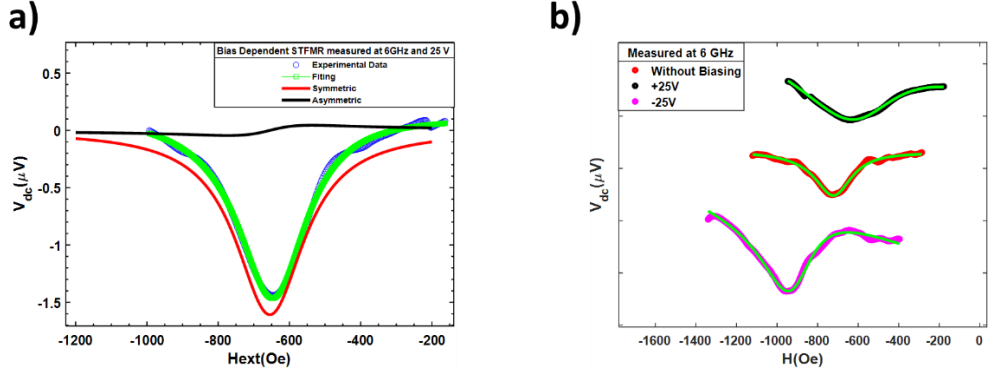

Figure 7. (a) Measured STFMR signal (green) and fitted signal (blue) with symmetric Lorentzian (red) and anti-symmetric Lorentzian (Black) as described by equation 2. (b) STFMR signal at 6 GHz in the presence of an external voltage of 0V,±25. Shift in H$_{reso}$ can be observed with and without voltage.

We measured a significant shift in the resonance field upon analyzing data and fitting it with Kittel's equation as given by Eq. (3). We obtained $H_\parallel$ 5 (Oe) as and $H_\perp$ as 6.9 (KOe).

$$f_{reso} = \frac{\gamma}{2\pi}\sqrt{(H_{ext} + H_\parallel)(H_{ext} + H_\parallel + H_\perp)} \qquad (3)$$

Upon obtaining values of $V_{sym}$, $V_{Asym}$ and $H_{reso}$ from the Eq. (2), we use Eq. (4) to calculate the effective spin Hall angle($\theta_{sh}$) for pristine and Pt-Porphyrin functionalized SLG devices, where $t_{py}$ and $t_{Gr}$ denote the thickness of Py and SLG layers respectively, and $M_s$ denote saturation magnetization of Py layer. By using $M_s = 8.36E5$ (A/m), $t_{Py} = 5$ nm, and $t_{Gr} = 1.2$ nm (obtained from AFM measurements), we calculated $H_\perp = 4.5$ (kOe) from Kittel's fit. After putting all numerical values in Eq. (2) we obtained the $\theta_{sh,Gr}$ to be 0.0013 and for Pt-Porphyrin functionalized SLG $\theta_{sh,\text{Pt-Porphyrin}}$ to be 0.016, which is one order larger than pristine Graphene.

This signifies that Pt-Porphyrin functionalized SLG converts charge current into spin current more efficiently than SLG because of the enhanced SOC at the interface. The $\theta_{sh}$ contributes to spin torque effects and is crucial for switching magnetic states. When combined with VCMA, torques exerted on the magnetization of Py upon absorbing spin current generated at the interface due to increased $\theta_{sh}$ can enhance the efficiency of magnetization switching at low-energy and fast-operating spintronics-based memory and logic devices.

$$\theta sh = \frac{V_{sym}}{V_{Asym}}\left(\frac{e}{\hbar}\right)\mu_0 M_s t_{Gr} t_{Py}\sqrt{1+\frac{Hper}{Hreso}} \qquad (4)$$

To further study the increase in SOC, we plotted the linewidth ($\Delta H$) vs $f_{RF}$ as shown in Fig. 8 and fitted with a linear Eq. (5). Where $\gamma$ is a gyromagnetic ratio, $\alpha_{eff}$ is the effective damping constant. It shall have an intrinsic and interface damping contribution, and $\Delta H_0$ is an Inhomogeneous linewidth broadening and is an intrinsic property of ferromagnetic thin film, and it gives a signature of the quality of the ferromagnetic thin film, the lesser the value better the quality. We obtained the value of $\Delta H_0$ 8(Oe). We calculated the slope of the linear fit and obtained a damping constant $\alpha_{eff,\text{PristineGr}}$ is 0.042 and $\alpha_{eff,\text{Pt-Porphryin}}$ is 0.056. A significant increase in damping in doped samples indicates enhanced spin relaxation processes due to the strengthened SOC at the interface. However, the values are in the ballpark with materials used in Magnetic tunnel junction (MTJ) devices and hence will have no deteriorating impact on switching speed. After confirming enhancement in SOC we then applied external potential through the back gate and measured the modulation of $\alpha_{eff,Pt\text{-}Porphryin}$ with external voltage as shown in Fig 8.(b).

$$\Delta H = \Delta H_0 + \frac{4\pi f \alpha_{eff}}{|\gamma|} \quad (5)$$

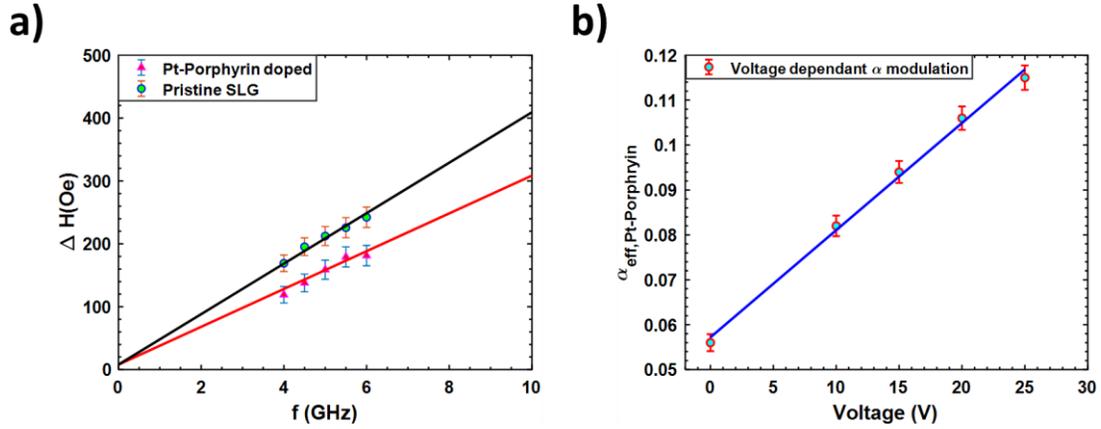

Figure 8. (a) Measured $\Delta H$ vs $f_{RF}$ for pristine graphene (red line) and Pt-Porphyrin doped SLG (black line). (b) Modulation of damping constant "$\alpha$" with an external back-gated voltage.

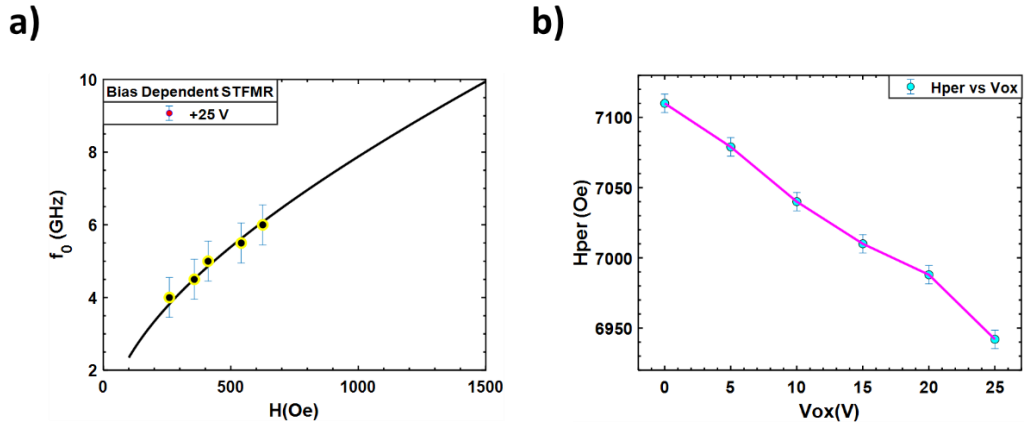

Figure 9 (a) Kittels fit for +25V external voltage. (b) Measured resonance field $H_{per}$ in the presence of external back gate voltage

The VCMA coefficient ($\xi$) is defined as the change in the surface anisotropy energy per unit electric field. To quantify, we assess the change in perpendicular field ($H_{per}$) with voltage, as illustrated in Fig. 9(b). We use Eq. (6) to calculate $\xi$, $M_s$ is the saturation magnetization and is 8.36e5 (A/m), $t_{Gr}$ is the thickness of SLG measured using AFM and is 1.2e-9(m), $t_{Py}$ is 5e-9(m), $H_\perp$ is 7.1 (KOe). It is a perpendicular magnetic anisotropy field at zero external potential, and we obtained this from Kittle's fit. We graphically represented $H_{per}$ against Voltage as illustrated in Fig. 11.

$$\xi = \frac{t_{FM} t_{ox} \mu_0 M_s}{2} \frac{dH_{anisotropy}(V)}{dV_{ox}} \quad (6)$$

$H_\perp$ will be regulated with an applied Electric field ($\vec{E}$) across the oxide and is given by $\vec{E} = \frac{V_{ox}}{t_{ox}}$ where $V_{ox}$ is the back gate DC voltage and $t_{ox}$ is the thickness of $SiO_2$. Therefore, we can write $H_{anisotropy}(V_{ox})$ as given by Eq (7), nonetheless, $H_\parallel$ will remain independent of $\vec{E}$.

$$H_{anisotropy}(V_{ox}) = H_\parallel + H_\perp(V_{ox}) \quad (7)$$

Taking into consideration of the VCMA effect, total anisotropy will be reduced, and hence we can rewrite $H_\perp(V_{ox})$ as: the $H_\perp(V_{ox}) = H_\perp(0) - H_{\perp,VCMA}(V_{ox}) \quad (8)$

where $H_\perp(0)$ is the perpendicular field measured at $V_{ox} = 0$

Therefore, we can write that upon putting this in Eq. (7) we get

$$H_{anisotropy}(V_{ox}) = H_\parallel + H_\perp(0) - H_{\perp,VCMA}(V_{ox}) \quad (9)$$

$$\xi = \frac{t_{FM} t_{ox} \mu_0 M_s}{2} \frac{[H_\perp(0) - H_{\perp,VCMA}(V_{ox})]}{[V_{ox,top} - V_{ox,bottom}]} \quad (10)$$

This $\frac{[H_\perp(0) - H_{\perp,VCMA}(V_{ox})]}{[V_{ox,top} - V_{ox,bottom}]}$ can be calculated from the slope of the $H_{per}$ vs $V_{ox}$ graph. The same change can also be noted from Kittles graph (Fig. 9(a)). Putting all the obtained values in Eq. (10) we get a value of $\xi = 375.6\ [fJV^{-1}m^{-1}]$.

**Conclusion**
This study illustrates the capability of Pt-Porphyrin functionalized graphene to realize substantial VCMA effects, facilitating its application in energy-efficient spintronic devices. The material properties warrant further investigation for the advancement of novel Spintronics-based memory and logic devices. Future endeavours will concentrate on enhancing the functionalization process and investigating additional suitable ferromagnetic materials with functionalized graphene exhibiting a significant VCMA effect. We anticipate additional advancements in characterization techniques for the detailed study of interactions.

## Materials and Methods

For the Fabrication of a back-gated device to demonstrate ST-FMR measurements, we selected an oxidized P-type (100) oriented silicon wafer with a resistivity of 0.05 Ω-cm, with a thermally grown 300nm dry oxide followed by foaming gas annealing. The selection of wafers was driven by two main factors (i) minimal substrate leakage current, and (ii) visually observing SLG on an oxidized silicon substrate under a microscope. We obtained SLG from Graphenea Inc., a CVD-grown SLG cultivated on copper foil coated with 2% Poly methyl methacrylate (PMMA). The polymer layer not only makes good visualization but also enables the effective transfer of SLG to the designated substrates while protecting the graphene surfaces from potential damage and contamination throughout the transfer process. A wet-etch technique was utilized to transfer SLG onto the oxidized silicon substrate. We began by etching copper with a copper etchant and then proceeded to clean the SLG/PMA in deionized (DI) water to remove copper etchant traces and we repeated the cleaning step 4-5 times so to remove any possible leftover traces. Following the cleaning process, the SLG/PMA was transferred to substrates using a self-designed and optimized scoop technique. After thoroughly drying the Substrate/SLG/PMA, we placed it in Acetone overnight to remove the PMA coating. We then rinsed the substrate in DI water followed by Rapid Thermal Annealing (RTP) in an argon atmosphere at 300°C to remove any residual PMMA traces. We purchased Platinum (II) 5,10,15,20-(tetraphenyl) porphyrin from Porphychem [26] and prepared a liquid solution with Molarity (M) 2.9e-4(mol/ltr) in Toluene. We also prepared M/4 and M/10 molarity solutions to study their impact on the intrinsic properties of pristine graphene. We then spin-coated the prepared solutions on the surface of SLG at 500rpm, followed by a heat treatment at 110°C under a hot wall inert gas flow oven. This step was conducted to augment surface reactions and facilitate π-π interactions between graphene and porphyrin [27]. Researchers have demonstrated the deposition of metalloporphyrins using the Atomic Layer Deposition (ALD) [28].

## Device Fabrication and Characterization

We followed the standard CMOS-compatible fabrication process flow for fabricating devices. We deposited a stack of film (Py (5nm)/Al (2nm)/SiO$_2$ (2nm)) on functionalized SLG and contact metal (Cr (10nm)/Au (250nm)) using the Sputtering tool "Orion" from the AJA international system. All depositions were done at a vacuum of 2e-3 (Torr) and base vacuum better than 5e-8 (Torr). We deposited the stack at a grazing angle without rotation. A capping layer stack (Al (2nm)/SiO$_2$ (2nm)) is used to avoid oxidation of Py. We then patterned SLG/stack into rectangular stripes of 20µm×80µm by positive photoresist AZ-5214E on an optical lithography mask aligner, followed by Ar-Ion milling for Etching. AZ-5214E resist doesn't dope the graphene; hence, edges will maintain intrinsic material properties like the channel. We removed the resist after etching using AZ 100 remover. For making electrical contacts on the fabricated devices, we patterned and deposited a stack of Cr (10nm)/Au (250nm) for contact to probe the device, followed by the Lift-off technique to remove metal from the field and inactive area on the wafer. We conducted ST-FMR measurements using a Bias-Tee, which allowed us to apply RF Power using a VNA (R&S model no. ZNB-20) at one of its capacitive ports and simultaneously measure the generated DC voltage at its inductive port. We measured the $V_{DC}$ signal with the help of a Nano-voltmeter (Keithley model no. 2182A). We did VNA Calibration using a CS-8 calibration substrate from GGB Industries. We also used high-frequency GS probes from the same company. We have measured in a frequency range from 4 to 7 GHz at a fixed power of +5dBm. We applied the external magnetic field at 45˚ to the external RF current and swept the external magnetic field from ± 2000Oe.

## Acknowledgement

We would like to acknowledge the support of the Indian Institute of Technology-Bombay Nanofabrication Facility (IITBNF) and the Industrial Research and Consultancy Center (IRCC) at IIT-Bombay Mumbai, India, for their financial support.

**Conflict of Interest**

The authors declare no conflict of interest.

**Author contribution:** ASS planned and designed experiments for transferring and functionalization of SLG and carried out the device fabrication with help from AE, DK, HAM, and AC. ASS measured the fabricated devices and carried out data analysis with the help of AAT. All authors commented on the manuscript. AAT planned and supervised the project

**Corresponding Author**

* Email: ambika.iitb14@gmail.com , ashwin@ee.iitb.ac.in